\documentclass[useAMS,usenatbib]{mn2e}
\usepackage{rotating}
\usepackage{journals}
\usepackage{graphicx, subfigure}
\usepackage{xspace}
\usepackage{amssymb}

\def\farcs{\hbox{$.\!\!^{\prime\prime}$}}
\def\degr{\hbox{$^\circ$}}
\def\arcmin{\hbox{$^\prime$}}
\def\arcsec{\hbox{$^{\prime\prime}$}\xspace}

\def\kev{$\rmn{keV}$\xspace}

\def\chan{{\it Chandra}\xspace}
\def\swift{{\it Swift}\xspace}
\def\rxte{{\it RXTE}\xspace}

\def\src{XTE~J1752-223\xspace}

\newcommand\msun {M$_{\odot}$\xspace}

\newcommand{\ergsec}{$\rmn{erg\,s^{-1}}$}

\LARGE \normalsize \title[\src: outburst decay and quiescence]{The black hole
  candidate \src towards and in quiescence: optical and simultaneous X--ray~--~radio observations}

\author[Ratti et al.]  {E.M.~Ratti$^{1}$\thanks{email : e.m.ratti@sron.nl}, P.G.~Jonker$^{1,2,3}$, J.C.A.~Miller--Jones$^{4}$,  M.A.P.~Torres$^{1,2}$,  J.~Homan$^{5}$,  \newauthor S.~Markoff$^{6}$, J.A.~Tomsick$^{7}$, P.~Kaaret$^{8}$, R.~Wijnands$^6$, E.~Gallo$^9$, \newauthor F.~$\ddot{\rmn{O}}$zel$^{10}$, D.T.H.~Steeghs$^{2,11}$ , R.P.~Fender$^{12}$  \\
$^1$SRON, Netherlands Institute for Space Research, Sorbonnelaan 2, 3584~CA, Utrecht, The Netherlands\\ 
$^2$Harvard--Smithsonian Center for Astrophysics, 60 Garden Street, Cambridge, MA~02138, U.S.A.\\
$^3$Department of Astrophysics, IMAPP, Radboud University Nijmegen, Heyendaalseweg 135,6525 AJ, Nijmegen, The Netherlands \\
$^4$International Centre for Radio Astronomy Research - Curtin University, GPO Box U1987, Perth, WA 6845, Australia \\
$^5$MIT Kavli Institute for Astrophysics and Space Research, 70 Vassar Street, Cambridge, MA 02139, U.S.A \\
$^6$Astronomical Institute ÔAnton PannekoekÕ, University of Amsterdam, P.O. Box 94249, 1090 GE Amsterdam, the Netherlands \\
$^7$Space Sciences Laboratory, 7 Gauss Way,ÊUniversity of California, Berkeley, CA 94720-7450, U.S.A \\
$^8$Department of Physics and Astronomy, University of Iowa, Iowa City, IA 52242, U.S.A \\
$^9$Department of Astronomy, University of Michigan, 500 Church St., Ann Arbor, MI 48109, U.S.A \\
$^{10}$University of Arizona, Department of Astronomy, 933 N. Cherry Ave., Tucson, AZ 85721, U.S.A \\
$^{11}$Department of Physics, University of Warwick, Coventry CV4 7AL, UK \\
$^{12}$ School of Physics and Astronomy, University of Southampton, Highfield, Southampton, SO17 1BJ, UK \\
}

\begin{document}

\maketitle

\begin{abstract} \noindent 

We present optical, X--ray and radio observations of the black hole transient (BHT) \src towards and in quiescence. Optical photometry shows that the quiescent magnitude of \src is fainter than 24.4 magnitudes  in the i$^\prime$-band. A comparison with measurements of the source during its 2009-2010 outburst shows that the outburst amplitude is more than 8 magnitudes in the i$^\prime$-band.
Known X--ray properties of the source combined with the faintness of the quiescence optical counterpart and the large outburst optical amplitude point towards a short orbital-period system (P$_{orb}\lesssim6.8\,{\rmn h}$) with an M type (or later) mass donor, at a distance of 3.5$\lesssim$d$\lesssim$8$\,{\rmn {kpc}}$. 
Simultaneous X--ray and radio data were collected with \chan and the EVLA, allowing constraints to be placed on the quiescent X--ray and radio flux of \src. 
Furthermore, using data covering the final stage of the outburst decay, we investigated the low luminosity end of the X--ray~--~radio correlation for this source and compared it with other BHTs. We found that \src adds to the number of outliers with respect to the `standard' X--ray~--~radio luminosity relation.  Furthermore, \src is the second source, after the BHT H1743$-$322, that shows a transition from the region of the outliers towards the `standard' correlation at low luminosity.
Finally, we report on a faint, variable X--ray source we discovered with \chan at an angular distance of $\sim$2\farcs9 to \src and at a position angle consistent with that of the radio jets previously observed from the BHT. We discuss the possibility that we detected X--ray emission associated with a jet from \src.
\end{abstract}

\begin{keywords} stars: individual (\src) --- 
accretion: accretion discs --- stars: binaries 
--- X--rays: binaries
\end{keywords}

\section{Introduction} 

\src was discovered as a transient source in the Galactic Centre region by the {\it Rossi X--ray Timing Explorer} (\rxte), on 2009 October 23 \citep{Mar09}. It was soon proposed to be a Galactic black hole candidate \citep{Mar09_2}, i.e., a binary system where a black hole (BH) is accreting matter from a companion star. Most black hole candidates are transient sources (black-hole transients, BHTs) that occasionally undergo outbursts. During outbursts these sources can show a characteristic evolution through various `states', defined on the basis of their strongly correlated spectral and variability properties (see, e.g., \citealt{Rem06}, \citealt{Bel10}). After the discovery, \src was monitored by the \rxte, \swift and {\it MAXI} satellites: the X--ray behaviour of the source during the outburst matched the typical phenomenological picture for BHTs, confirming \src as strong accreting black hole candidate (\citealt{Nak10}, \citealt{Mun10}, \citealt{Shap10}, \citealt{Cur11}).  Based on the X--ray spectral and timing properties of \src, \citet{Shap10} also report a mass estimate for the BH, M$_{BH}=9.8\pm0.9$\,\msun and the distance to the source $d=3.5\pm0.4\,{\rmn {kpc}}$ (although the systematic uncertainties of these estimates could be large). 
\newline
A bright optical counterpart to \src was identified by \citet{Tor09} based on the \swift position  and later confirmed through optical spectroscopy by \citet{Tor09_2}.  Two radio sources were detected at a position consistent with the optical one \citep{Bro10} which were initially interpreted as a decelerated jet and its receding counterpart \citep{Yan10}. A combination of radio and optical observations with accurate astrometry allowed \citet{MiJ11} to locate the radio core of the source, at R.A. =$17^{\rmn{h}}$\,$52^{\rmn{m}}$\,$15\fs09509(2)$, Dec. = $-22\degr$\,$20\arcmin$\,$32\farcs 3591(8)$. This position lies to the southeast of the two jet components previously observed, which were then re-interpreted as two ejection events. The core position was recently confirmed by \citet{Yan11}, who also report on the radio detection of a third jet component. \newline
 After transiting through all the canonical states of a BHT, \src faded towards quiescence in July 2010 \citep{Rus12}.
The quiescent state of BHTs was initially considered as an extension towards low luminosities (10$^{30}$-10$^{33}$ \ergsec) of the `hard state', i.e. a spectral state where the X--ray spectrum is dominated by a power-law component with index $\Gamma\sim1.5$. Later, a number of BHTs were found to show softer spectra in quiescence compared to the hard state, suggesting that the former can be considered a state on its own (e.g., \citealt{Jon04}, \citealt{Tom04}). Still, because of their low luminosities at all wavelengths, it has been challenging to collect high quality spectral data to constrain the properties of quiescent BHTs. 
\begin{figure}
\includegraphics[width=10.0cm, angle=-90]{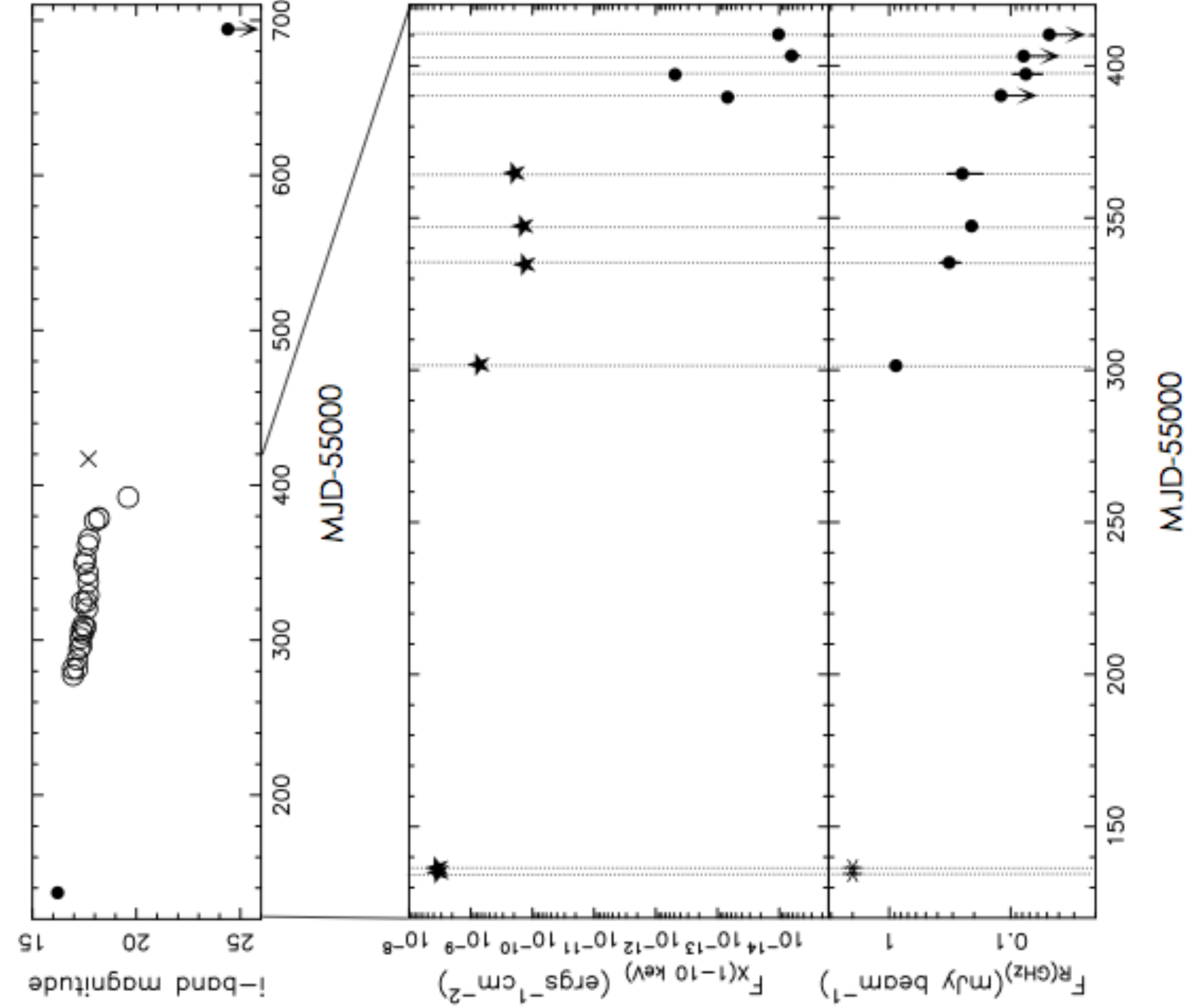} 
\caption{Top panel: i$^\prime$  band light curve from the optical counterpart to \src, including our IMACS observations (black dots), and the measurements obtained by \citet{Rus12} with the Faulkes telescopes (empty circles) and by \citet{Cor10_1} with the WHT/ACAM  (cross),  showing the \src optical rebrightening. Central panel: X--ray light curve from our \chan observations (black dots) and RXTE/PCA observations from \citet{Rus12} (stars). The 3-20 $keV$ count rate provided by the authors has been converted into 1-10 $keV$ fluxes using the HEASARC tool {\sc WebPIMMS}, assuming a power-law spectrum with a photon index of 1.6 or 1.4 (spectral information from \citealt{Rus12} and \citealt{Shap10}). Bottom panel: GHz radio light curve including our EVLA and VLBA observations (black dots) and the hard-state observations reported by \citet{Bro09}. The dotted lines across the radio and X--ray light curves highlight the temporal correspondence of the data in the two energy bands. Note that the X--ray and radio light curves cover a shorter time range than the i$^\prime$-band light curve in the top panel.}
\label{licus}
\end{figure}
\begin{table*}
\caption{ A journal of the \chan observations.}
\label{log}
\begin{center}
\begin{tabular}{cccccc}
\hline
Obs ID & Observing & MJD & Time on      & Count rate                         & Detected \#source counts \\
             & date          & (days; UTC) & source (ks) & 0.3-7 keV (cnt s$^{-1}$) & 0.3-7 keV  \\
\hline
\hline
11053 & 2010 Jul.~12 & 55389.63922 & 6.36  & $(5.5\pm1.1)\times10^{-3}$ & 44 \\
12310 & 2010 Jul.~20 & 55397.07034 & 13.6 & $(3.9\pm0.2)\times10^{-2}$& 548 \\
11055 & 2010 Jul.~26 & 55403.24697 & 31.4  & $(6.2\pm1.6)\times10^{-4}$& 20 \\
11056 & 2010 Aug.~02& 55410.27429& 88.9  & $(8.7\pm1.0)\times10^{-4}$ & 79\\
\hline
\end{tabular}		      
\end{center}
\end{table*}
\newline
The hard state is also associated with the presence of radio emission, with a flat or slightly inverted spectrum that, owing to the high brightness temperature, is generally thought to originate from a compact jet. The X--ray emitting accretion flow and the radio emitting jet are known to be intimately connected:  \citet{Gal03} and \citet{Corb03} found that several BHTs follow a correlation between the X--ray and radio luminosity (L$_X$ and L$_R$ respectively) in the form L$_R\propto$L$_X^{0.7}$. The power-law index was later refined to L$_R\propto$L$_X^{0.6}$ \citep{Gal06}. The correlation was initially thought to be universally valid for all BHTs, but in recent years a number of outliers have been found (see \citealt{Cal10} for an updated compilation of sources). For most of the outliers it is not established yet whether they follow the correlation at a lower normalization or a correlation with a different slope. The work of \citet{jon10} and \citet{Cor11} on the outlier H1743$-$322 has shown that a reconnection of the  the `outliers branch' with the standard correlation is also possible:  H1743$-$322 in fact lies on a steeper correlation than L$_R\propto$L$_X^{0.6}$ at high luminosity \citep{jon10}, but undergoes a transition back to the canonical correlation as the luminosity decreases below $\sim10^{34}$\,\ergsec \citep{Cor11}.  
This indicates a change-over between two accretion regimes when the source moves from a high-luminosity to a low-luminosity hard state. Following the X--ray~--~radio correlation across a broad range in luminosity for other BHTs may show whether similar transitions are a common feature among this class of sources and provide new elements to our understanding of the accretion mechanism at different accretion rates, including in the low-luminosity quiescent regime. It may also help us understand why different sources follow different X--ray~--~radio correlations.
\newline
Here, we report on contemporaneous \chan X--ray and Expanded Very Large Array (EVLA) radio observations of \src aimed at following the X--ray and radio light curves and establishing the X--ray~--~radio correlation during the final part of the decay towards quiescence. We also present one Very Long Baseline Array (VLBA) detection and optical observations of \src in quiescence, providing information about the system's orbital period and distance. 

\section{Observations, data reduction and results}

\subsection{{\it Chandra} X--ray observations} 

\begin{figure}
\includegraphics[width=8.5cm, angle=0]{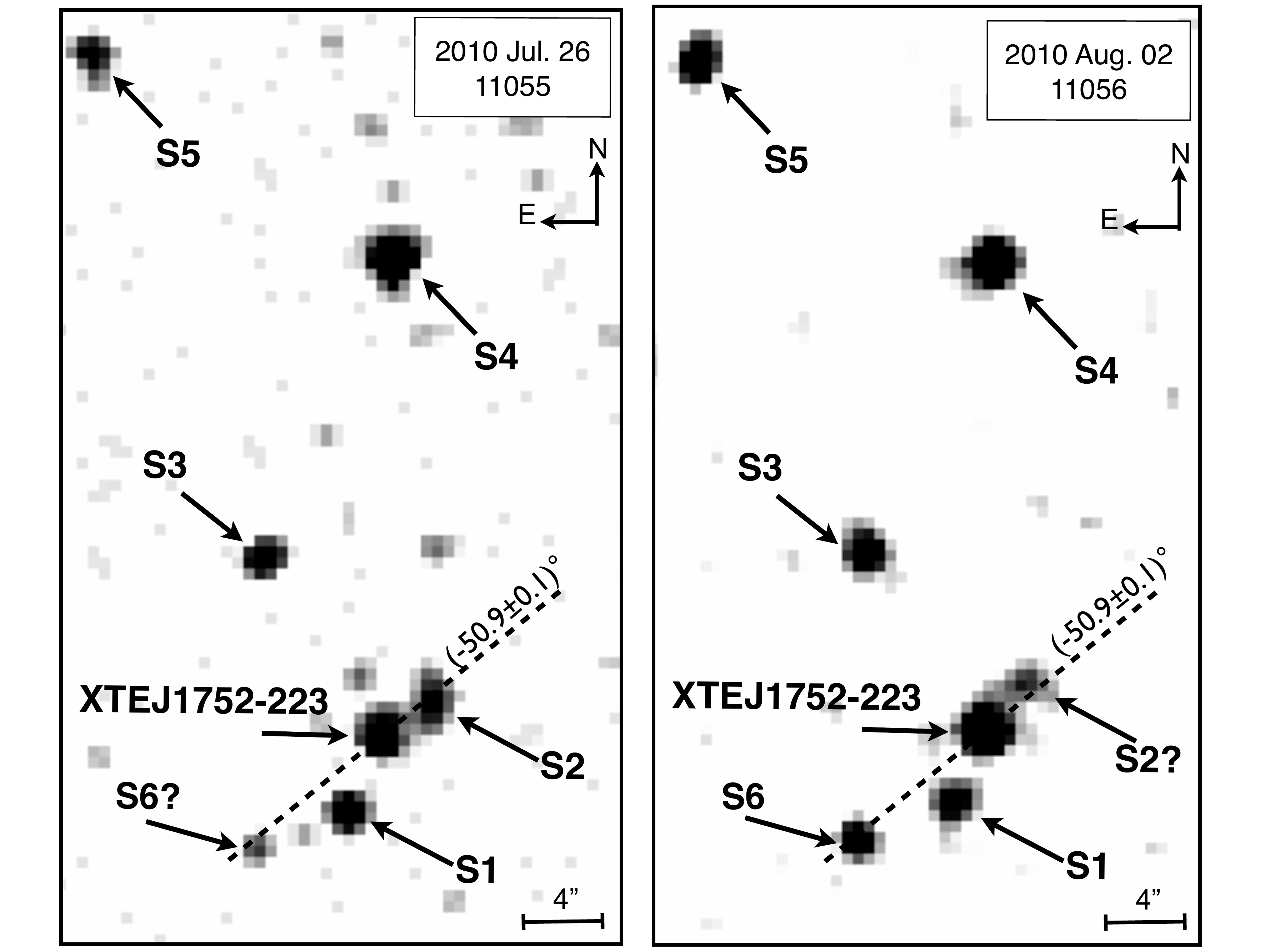} 
\caption{0.3-7 $\rmn{keV}$ \chan/ACIS-S images from observation 11035 and 11036 ({\it left panel} and {\it right panel} respectively). The data are Gaussian smoothed with a kernel radius of 2. The finding charts show 6 unidentified sources detected by \chan close to \src (see also Table \ref{tab:csources}), labeled S1 to S6 (with a question mark if the detection is uncertain). The dashed line shows the position angle of the jet component named A by \citep{Yan10} at the time of its last detection, with respect to \src (-50.9$\pm$0.1 degrees, \citealt{MiJ11}). This is consistent with the position angle of S2 (-52$\pm$7 degrees) with respect to \src.}
\label{chanfinders} 
\end{figure}

We observed \src with the \chan satellite using the
back--illuminated S3 CCD--chip of the Advanced CCD Imaging
Spectrometer (ACIS) detector (\citealt{Gar97}) on
four occasions during the decay towards quiescence (see Figure \ref{licus} and Table~\ref{log}).  During all the observations the ACIS--S3 detector was
windowed, providing a frame time of 0.4104~s.
We have reprocessed and analysed the data using the {\sc CIAO}
software developed by the \chan X--ray Center\footnote{http://cxc.cfa.harvard.edu/ciao4.3/}. Since, by design, the source position falls near the optical
axis of the telescope, the size of the point spread function is smaller
than the ACIS pixel size. Therefore, we follow the method of
\citet{Li04} implemented in the {\sc CIAO 4.3} tool
{\sc acis\_process\_events} to improve the image quality of the ACIS
data.
In our analysis we
have selected events only if their energy falls in the 0.3--7 keV
range.  All data have been used, as background flaring is very weak or
absent in all data. The last observation  (ID 11056) has been performed
with the datamode set to {\sc VFAINT}. This means that pulse height
information in a 5x5 pixel region around the event is telemetered
down, allowing for a more rigorous cleaning of background events
caused by for instance cosmic rays. 
\newline
Using {\sl wavdetect} we detected \src in each of the observations. We
selected a circular region of 10\arcsec\,radius centred on the
accurately known source position \citep{MiJ11} to extract
the source counts for the first two observations (Obs IDs 11053 and
12310).  The longer, deeper, exposures of the last two observations
revealed faint sources near the position of \src (Figure \ref{chanfinders}). Therefore, we used a smaller extraction radius of 1\farcs5. 
In order to correct the source flux in these latter two observations for the small extraction radius, we used the {\sc arfcorr} command in {\sc CIAO}. 
For all four \chan observations, we used a circular
region with a radius of 10\arcsec\,on a source-free region of the CCD
to extract background counts. The redistribution response file is the
same for the source and background region but we have made auxilliary
response matrices for the source region of each of the observations
separately. The net, background subtracted, source count rate for each
observation is given in Table~\ref{log}.
\newline
Using {\sc xspec} version 12.4.0ad (\citealt{Arn96}) we fit
the spectra of \src using Cash statistics
(\citealt{Cas79}) modified to account for the
subtraction of background counts, the so called
W--statistics\footnote{see
  http://heasarc.gsfc.nasa.gov/docs/xanadu/xspec/manual/} for all four
observations. We used an absorbed power--law model ({\sl
  pegpwrlw} in {\sc xspec}) to describe the data.
\newline
Due to the relatively low number of detected counts, we fixed the interstellar
extinction during the fits to 5$\times 10^{21}$ cm$^{-2}$ found by
\citet{Cur11}. The power-law index and normalisation
were allowed to float. The results of our spectral analysis are listed in Table~\ref{spec}.
We note that the {\sc goodness} percentage is in all cases far from the nominal 50\%. However, by visually inspecting the spectra there is no clear reason to reject the fits. Possibly the low number of counts is responsible for the discrepant goodness values.

\begin{table}
\caption{Best fit parameters of the X--ray spectra of \src. PL refers to
  power law.  All quoted errors are at the 68 percent confidence
  level. N$_H$ was fixed in all instances to 5$\times 10^{21}$ cm$^{-2}$. }
\label{spec}
\begin{center}
\begin{tabular}{ccccc}
\hline
Obs ID  & PL index & Unabs.~0.5--10 keV flux & Goodness \\ 
&    &  erg$^{-1}$~cm$^{-2}$~s$^{-1}$  & percent \\
\hline
\hline
11053 &  1.6$\pm$0.5 & $(7.0\pm2.0)\times 10^{-14}$  & 0\\
12310 &  1.6$\pm$0.1 & $(6.4\pm0.6)\times 10^{-13}$  & 91 \\
11055 &  1.9$\pm$0.5 &  $(1.0\pm0.2)\times 10^{-14}$  & 20\\
11056 &  1.7$\pm$0.2 &  $(1.4\pm0.2)\times 10^{-14}$  & 11\\
55$+$56$^a$ & 1.8$\pm0.2$  & ($1.3\pm0.1)\times 10^{-14}$ & 10 \\
\hline
\end{tabular} \\
\footnotesize{$^a$ Fit using the spectra of 11055 and 11056 combined.} \\
\end{center}
\end{table}

\subsection{Quiescent X--ray emission from \src}
\label{sec:xres}
The X--ray spectra we acquired are dominated by a power-law component, indicating that the source was in a hard spectral state at the time of the observations. As shown in Table \ref{spec}, the  power-law index $\Gamma$ is $\sim$1.6 in the first two observations (11053 and 12310), while the last two observations (11055 and 11056) present a slightly softer spectrum (although $\Gamma$ is consistent with a constant value across the observations at the 1$\sigma$ level). 
The light  curve in Figure \ref{licus} shows the unabsorbed X--ray flux from \src,  from our \chan observations at the beginning of the quiescent phase and from RXTE observations performed earlier in the outburst. The unabsorbed flux on 2010 Jul. 12 was the faintest observed from this source since the beginning of the outburst, (7.0$\pm$2.0)$\times$10$^{-14}$ \ergsec\,in the 0.5-10\,$\rmn{keV}$ range. In the following week \src experienced a re-brightening by (nearly) one order of magnitude, reaching (6.4$\pm$0.6)$\times$10$^{-13}$ \ergsec\,on  2010 Jul. 20. Six days later, on Jul. 26, the source had faded again down to (1.0$\pm$0.2)$\times$10$^{-14}$ \ergsec. The flux level at the time of our last observation, on 2010 Aug. 02, is consistent (within 2$\sigma$) with that shown on Jul. 26 (see Table \ref{spec}), suggesting a flattening of the light curve. Despite the large errorbars, the flux level in the last two observations is in fact inconsistent with a constant decay after the 2010 Jul. 20 flare. This supports the conclusion that the source has reached quiescence (see discussion in Section \ref{sec:disc}).  \newline

\begin{table*}
\caption{New \chan sources detected in the vicinity of \src and their candidate counterparts in the i$^\prime$-band. Labels in column 1 refer to Figure \ref{chanfinders}. The uncertainty on the X--ray positions is dominated by the 0.6 \arcsec boresight error of \chan. The count rate is that measured by {\sc wavdetect} on the deepest \chan image, 11056,  for all the sources but S2. For the latter, the counts are measured on the image 11055. The last two columns report the position of optical sources detected within the \chan error circle (see text in Section \ref{coun}). The accuracy on the optical position is 0\farcs05 on both  R.A. and Dec. }
\label{tab:csources}
\begin{center}
\begin{tabular}{lcccccc}
\hline
Label & \chan name & R.A & Dec. & Count rate (cnt s$^{-1}$) & R.A. & Dec.  \\ 
&  & (\chan) & (\chan)  &  & (i$^\prime$-band) & (i$^\prime$-band)  \\ 
\hline
\hline
S1& CXOU J175215.1-222035  & $17^{\rmn{h}} 52^{\rmn{m}} 15\fs19$ &  $-22\degr 20\arcmin 35\farcs4$  & (2.6$\pm$0.5)$\times$10$^{-4}$ 
 & $17^{\rmn{h}} 52^{\rmn{m}} 15\fs2$ &  $-22\degr 20\arcmin 34\farcs5$ \\
S2 & CXOU J175214.8-222030    &  $17^{\rmn{h}} 52^{\rmn{m}} 14\fs91$ &  $-22\degr 20\arcmin 31\farcs4$ & (5.8$\pm$1.3)$\times$10$^{-4}$ 
& $-$ & $-$ \\
S3 & CXOU J175215.4-222023 &  $17^{\rmn{h}} 52^{\rmn{m}} 15\fs46$ &  $-22\degr 20\arcmin 23\farcs9$  & (5.6$\pm$0.8)$\times$10$^{-4}$ 
 & $17^{\rmn{h}} 52^{\rmn{m}} 15\fs42$ &  $-22\degr 20\arcmin 24\farcs 3$ \\
 &    &   &  &  
 & $17^{\rmn{h}} 52^{\rmn{m}} 15\fs45$ &  $-22\degr 20\arcmin 24\farcs 5$ \\
S4 & CXOU J175215.0-222010 &  $17^{\rmn{h}} 52^{\rmn{m}} 15\fs06$ &  $-22\degr 20\arcmin 10\farcs9$ & (8.8$\pm$1.0)$\times$10$^{-4}$ 
 & $17^{\rmn{h}} 52^{\rmn{m}} 15\fs10$ &  $-22\degr 20\arcmin 11\farcs 4$ \\
 &    &   &  &  
 & $17^{\rmn{h}} 52^{\rmn{m}} 15\fs05$ &  $-22\degr 20\arcmin 11\farcs 0$ \\
S5 & CXOU J175215.9-222001 &  $17^{\rmn{h}} 52^{\rmn{m}} 16\fs01$ &  $-22\degr 20\arcmin 02\farcs25$ & (4.3$\pm$0.7)$\times$10$^{-4}$ 
 & $17^{\rmn{h}} 52^{\rmn{m}} 16\fs02$ &  $-22\degr 20\arcmin 01\farcs 5$ \\
S6 &  CXOU J175215.4-222036 &  $17^{\rmn{h}} 52^{\rmn{m}} 15\fs49$ &  $-22\degr 20\arcmin 37\farcs0$ & (2.1$\pm$0.5)$\times$10$^{-4}$ 
 & $17^{\rmn{h}} 52^{\rmn{m}} 15\fs50$ &  $-22\degr 20\arcmin 37\farcs2$ \\
\hline
\end{tabular} 
\end{center}
\end{table*}

\subsection{New \chan sources in the vicinity of \src}
Figure \ref{chanfinders} shows the two deepest images we acquired with \chan, 11055 and 11056 (see Table \ref{log}). Six unidentified sources are detected in the vicinity of \src, labeled S1 to S6 (see also Table \ref{tab:csources}). The closest to \src is CXOU J175214.8-222030 (S2), detected only in the observation 11055 with 10 net counts at  an angular separation of $\sim$2\farcs9 from \src. The source position provided by {\sc wavdetect} defines a position angle with respect to \src of -52$\pm$7 degrees. The significance on the flux measurement from {\sc wavdetect} is 4.1 $\sigma$. In the 0.3-7\,\kev band, the probability of finding the source by chance is less than 3$\times$10$^{-15}$, corresponding to more than 8 $\sigma$ in Gaussian statistics. S2 is not significantly detected by {\sc wavdetect} in any other of our \chan observations, although visual inspection  of  the  deepest one, 11056, shows a faint source close to the position of S2 in 11055.  Considering a 1\arcsec radius circle  (\chan $\sim$95\% encircled energy radius)  centred by eye on this source provides $\sim10$ counts in the 03-7\,\kev band. The estimate of the background is such that, from Poisson statistic only, the probability that the source is due to a statistical fluctuation of the background is less than $10^{-14}$. 
By considered the same 1\arcsec radius circle, we calculated 95\% confidence upper limits to the count rate of S2 in all the observations where it was not detected. 
The upper limits are summarized in Table \ref{tab:UL}.  Count rates were transformed into unabsorbed fluxes by assuming N$_H=5\times10^{21}\,\rmn{cm}^{-2}$  and a power-law spectrum with a photon index of 1.6 (similarly to the X--ray spectrum of the jet feature observed by \citealt{Corb05} in H1743-322).  \newline
The source CXOU J175215.4-222036 (S6) is also detected only once, in the deepest observation 11056. The 4 counts collected in 11055 at the source position do not provide a significant detection, as there is a Poissonian probability of 13\% to detect as many counts from S6  if the source was at the same flux level as in 11056. 

\subsection{Optical observations}

\begin{figure*}
\includegraphics[width=16.0cm, angle=0]{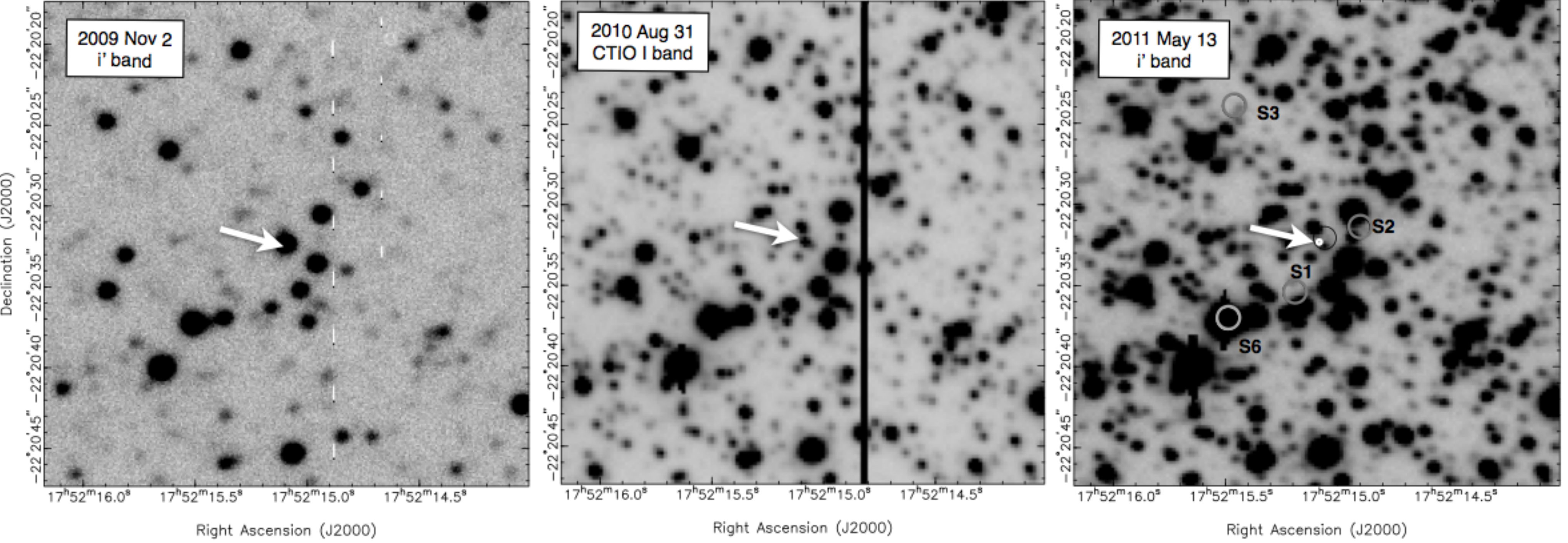} 
\caption{Finding charts: the position of \src (from radio observations in \citealt{MiJ11}) is indicated by the arrow and further highlighted by the white dot in the rightmost finding chart, where no counterpart to \src is detected. The error-circles in the same finding chart indicate the position of nearby unidentified \chan X--ray sources,  labeled as in Figure \ref{chanfinders}. As we are using our astrometric solution in order to over-plot the X--ray positions on the finding charts, the error-circles in the figure account for the accuracy of \chan and of our astrometry on the optical images, added linearly. The resulting 90\% confidence radius is 0\farcs68. The \chan position of \src is indicated by the thin black error-circle. The black band on the central finding chart is caused by a bad column on the IMACS CCDs.}
\label{finders}
\end{figure*}

We collected three observations of \src in the i$^\prime$  and I (CTIO) bands, with the Inamori-Magellan Areal Camera and Spectrograph (IMACS) at the Magellan Baade telescope in Cerro Las Campanas on 2009 Nov.~  2 (5 s exposure, seeing 0\farcs8), 2010 Aug.~31 (3 exposures of 180 s each, seeing 1\farcs3) and 2011 May~13 (300\,s-long deep exposure plus 5\,s-long exposure for astrometry, seeing 0\farcs7). The images were corrected for bias and flat-fielded with standard routines running in {\sc midas} \footnote{http://www.eso.org/sci/software/esomidas/}. \newline
An astrometric solution was obtained (using {\sc midas}) against entries from the third U.S. Naval Observatory CCD Astrograph Catalog (UCAC3; \citealt{Zac10}), considering only sources that are not saturated on the CCD and appear stellar and unblended. The astrometric solution was fitted for the reference point position, the scale and the position angle, obtaining root-mean-square (rms)  residuals of $\sim$0\farcs03 for the 2009 Nov.~2 and 2011 May~13 images, and 0\farcs05 for the 2010 Aug.~31 exposure. The 300\,s exposure on  2011 May~13  was astrometrically calibrated using the 5\,s one as a secondary catalogue. The final astrometric solution was calculated on stars with UCAC3 fit model magnitudes in the 14-16.5 range, for which the positional accuracy of UCAC3 is estimated to be  0\farcs01 \citep{Zac10}. In addition, the systematic uncertainty in tying the UCAC3 stars to the International Celestial Reference System (ICRS) is 0.005\arcsec \citep{Zac10}. For the accuracy of our stellar positions we adopt the linear sum of the residuals of the astrometry and the accuracy of the catalogue (as the latter is potentially a systematic error): the resulting positional accuracy at 1$\sigma$ across different observations ranges between 0\farcs046 and 0\farcs065 on both right ascension and declination.
\newline
The photometry was performed through point spread function (psf) fitting, using {\sc daophot~II} \citep{Ste87} running within {\sc midas}. The absolute photometry of three comparison stars labeled C1, C2, C3 in \citet{MiJ11} (with i$^\prime$-band magnitudes of 13.62, 14.53 and 14.38 respectively) was used to determine the i$^\prime$-band photometric calibration. The  photometry of the 2010 Aug.~31 image  could not be calibrated due to the lack of observations of photometric standard stars in the I (CTIO) filter.

\begin{table}
\caption{ Constraints on the flux from CXOU J175214.8-222030 (S2 in Figure \ref{chanfinders}). The count rate is calculated within 1\arcsec ( $\sim$95\% encircled energy radius) around the position of S2. Upper limits are at the 95\% confidence level. A power-law spectrum with photon index 1.6 is assumed (see text), and N$_H=5\times10^{21}\, \rmn{cm}^{-2}$. }
\label{tab:UL}
\begin{center}
\begin{tabular}{ccc}
\hline
Obs ID & Observing Date & Unabs. 0.3-7 $\rmn{keV}$ flux  \\
             &                               &  (\ergsec$\rmn{cm}^{-2}$) \\
\hline
\hline
11053 & 2010 Jul. 12 & $<$17.8$\times10^{-15}$ \\
12310 & 2010 Jul. 20 & $<$5.1$\times10^{-15}$ \\
11055 & 2010 Jul. 26 & 4.6$\times10^{-15}$ \\
11056 & 2010 Aug. 2 & $<$2.6$\times10^{-15}$ \\
\hline
\end{tabular} 
\end{center}
\end{table}

\subsection{Optical outburst amplitude of more than 8 magnitudes}
\label{sec:optres}
Figure \ref{finders} shows 30\hbox{$^{\prime\prime}$}$\times$30\arcsec finding charts from our i$^\prime$ and I-band observations, where the optical counterpart to the radio core of \src (R.A.$\,=17^{\rmn{h}}\,52^{\rmn{m}}\,15\fs09509\pm0.00002$, Dec.$\,=-22\degr\,20\arcmin\,32\farcs3591\pm0.0008$, \citealt{MiJ11}) is indicated by arrows. The first observation, from 2009 Nov.~ 2 (Figure \ref{finders}, {\it left panel}), was taken during the first part of the 2009 X--ray outburst, when the source was in a low-hard spectral state (\citealt{Nak10}, \citealt{Mun10}). Accurate astrometry of this image has been used by \citet{MiJ11} in order to locate the radio core of \src.  The i$^\prime$-band magnitude of \src measured with our psf photometry in this observation is 16.29$\pm$0.01, consistent with the measurement obtained from aperture photometry by \citet{MiJ11}. The second observation  (Figure \ref{finders}, {\it middle panel}) was performed on 2010 Aug.~31, and shows the fading of the source towards quiescence, after the optical re-brightening occurred on 2010 Aug.~8 (\citealt{Cor10_1}, \citealt{Cor10_2}). In the third observation (Figure \ref{finders}, {\it right panel}), taken on  2011 May~13 after almost one year of quiescence, the optical counterpart to \src it is not detected anymore down to a limiting magnitude of 24.4 (3 $\sigma$ upper limit) in the i$^\prime$  filter.  Close neighbours to the optical counterpart are visible in the last two finders. One of them is within the \chan error circle, at $\sim$0\farcs4 to the X--ray position. Nonetheless, the association with \src is ruled out by the radio position and by the variability observed from the actual optical counterpart to the source. The psf photometry is able to resolve the counterpart to \src from those nearby stars in the observation taken during outburst, when those are outshone by the target's light. 
A comparison of the magnitudes measured from our first and last observations shows that the drop in magnitude from outburst to quiescence is more than 8 magnitudes. 

\subsection{Optical counterparts to unidentified \chan sources}
\label{coun}
\begin{figure}
\includegraphics[width=8.0cm, angle=0]{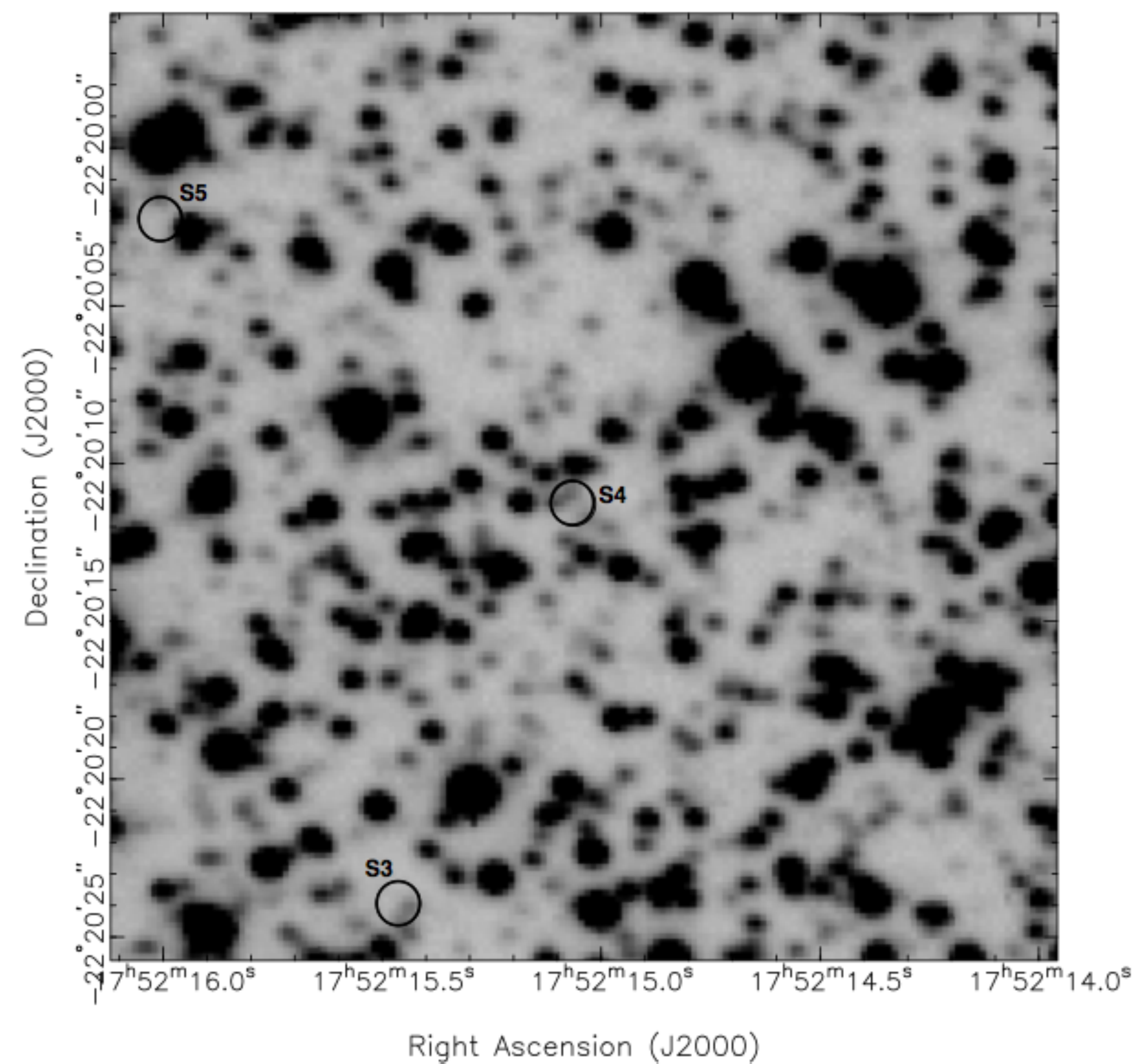} 
\caption{ \chan position of the unidentified sources labeled S3, S4 and S5 in Figure \ref{chanfinders} and Table \ref{tab:csources}, over-plotted to our IMACS i$^\prime$-band observation of 2011 May 13.  The error-circles have a 0\farcs7 radius, accounting for the \chan boresight and the accuracy of our astrometry on the optical images. }
\label{finderS4S5}
\end{figure}

The 2011 May~13 finding chart in Figure \ref{finders} shows the position of some of the unidentified X--ray sources detected by \chan during our observations ((S1, S2, S3, S6, see Section \ref{sec:xres}). Other two unidentified sources nearby (S4, S5) are shown in Figure \ref{finderS4S5}. There is no clear i$^\prime$-band counterpart to the faint X--ray source S2 detected close to \src, although there is a star near the edge of the 90\% \chan error-circle (0\farcs6 radius), at 0\farcs7 from the X--ray position (the 90\% uncertainty on the optical position is 0\farcs08). A brighter star is located a bit further, at $\sim$1\farcs10. Unfortunately, none of our optical images is close in time to the \chan observation 11055, where this source was brightest in the X--rays. For both S1 and S5, a faint source lies on the edge of the \chan error-circle. S6 can be associated with a bright star, while S3 has two faint optical counterpart candidates, partly blended together. Two faint optical sources are also consistent with the position of S4. The position of the best candidate counterpart(s)  to each \chan source is reported in Table \ref{tab:csources}.

\subsection{Radio observations: EVLA}
\label{sec:evla}
The new Expanded Very Large Array \citep[EVLA;][]{Per09} was used to monitor the
decay of the outburst of \src from 2010 April 15 through 2010 August
2, under program codes AM1039 and SB0329.  With the newly-operational wideband
4--8\,GHz receiver system, we were able to observe simultaneously in two
independent 128-MHz sub-bands (each comprising 64 channels of width 2\,MHz) to
obtain spectral information at every epoch.  To avoid the radio frequency
interference (RFI) known to exist below 4.5\,GHz, and yet achieve the widest
feasible frequency separation, the two sub-bands were centred at 4.6 and
7.9\,GHz. Once the source was no longer detected
in an individual sub-band, no spectral information could be derived.  We then
switched the frequency setup in order to achieve the maximum possible sensitivity, observing
over a contiguous bandwidth of  256\,MHz centred at 8.4\,GHz. Throughout our
observing campaign, the array was in its most compact `D' configuration, with an
angular resolution at frequency $\nu$ of 12\arcsec(6\,GHz$/\nu$). \newline
Data reduction was carried out using the Common Astronomy Software Application
\citep[{\sc CASA};][]{McM07}.  The data were initially averaged down by a factor of 10
from the default 1\,s integration time to make the data sets more manageable.
Baseline corrections were performed and bad data arising from shadowing,
instrumental issues, or RFI were edited out before beginning the calibration. 
Bandpass and flux density calibration was carried out using 3C\,286, setting the
flux scale according to the coefficients derived at the VLA in 1999
\citep{Per99}.  Amplitude and phase gains were derived for all calibrator
sources, referencing the target source XTE J1752--223 to the calibrator
J1820--2528.  OQ\,208 was used as an unpolarized calibrator to derive the
polarization leakage terms and 3C\,286 was used to calibrate the polarization
position angle.  Finally, the calibration was applied to the target source,
which, following frequency-averaging by a factor of 8, was then subjected to
several rounds of imaging and self-calibration.  Measured flux densities for XTE
J1752--223 are given in Table~\ref{tab:evla}.

\begin{table*}
\begin{center}
{\caption{\label{tab:evla} A journal of the EVLA observations}}
\begin{tabular}{llccccc}
\hline\hline
Observation date & MJ$^1$           & Frequency & Bandwidth & Flux density & Spectral index $\alpha$ \\
                 &(days; UTC)          &  (GHz)   & (MHz)     & (mJy\,beam$^{-1}$\\
\hline
\hline
2010 Apr 15   & $55301.43\pm0.01$ & 4.6 & 128 & $1.33\pm0.04$ & -0.7$\pm$0.1\\
                          & $55301.43\pm0.01$ & 7.9 & 128 & $0.88\pm0.04$& \\
2010 May 19   & $55335.29\pm0.01$ & 4.6 & 128 & $0.43\pm0.05$& -0.5$\pm$0.4\\
                          & $55335.29\pm0.01$ & 7.9 & 128 & $0.32\pm0.06$& \\
2010 May 31   & $55347.31\pm0.02$ & 4.6 & 128 & $0.18\pm0.03$& -0.3$\pm$0.4\\
                          & $55347.31\pm0.02$ & 7.9 & 128 & $0.21\pm0.02$&\\
2010 Jul.13   & $55390.17\pm0.02$ & 4.6 & 128 & $<0.10$& -\\
                        & $55390.17\pm0.02$ & 7.9 & 128 & $<0.12$& -\\
2010 Jul.20   & $55397.23\pm0.10$ & 8.4 & 256 & $0.075\pm0.020$&-\\
2010 Jul.26   & $55403.16\pm0.07$ & 8.4 & 256 & $<0.078$&-\\
2010 Aug.02   & $55410.20\pm0.10$ & 8.4 & 256 & $<0.048$&-\\
\hline
\end{tabular}
\end{center}
$^1$ the MJD is at the mid-point of the observation. The errorbar reflects the observation length. \newline
$^2$ upper limits are on the 3$\sigma$ level.
\end{table*}

\subsection{Radio observations: VLBA}
One epoch of VLBA data was also taken during the decaying hard state of XTE J1752--223, on 2010 June 17, under program code BM346.  We observed with nine VLBA antennas (the Pie Town antenna was out of the array owing to a broken rail).  We observed at 8.4\,GHz in dual circular polarization, with the maximum available recording rate of 512\,Mbps, corresponding to 64\,MHz of observing bandwidth per polarization.  The observations were phase referenced to the nearby calibrator source J1755--2232, from the third extension to the VLBA Calibrator Survey \citep[VCS-3;]{Pet05} and located $0.76^{\circ}$ from XTE J1752--223.  We switched between target and calibrator with a cycle time of 3\,min, substituting the VCS-5 \citep{Kov07} check source J1751--1950 for every eighth scan on the target. By observing a range of bright calibrator sources at differing elevations, for 30\,min at the start and end of the observing run (aka geodetic blocks), we could better solve for unmodeled clock and tropospheric phase errors. The geodetic blocks were analysed using the Astronomical Images Processin System ({\sc AIPS})\footnote{http://www.aips.nrao.edu/index.shtml} task {\it DELZN}, thereby improving the success of the phase transfer.  Data reduction was carried out according to standard procedures within {\sc AIPS}.  The flux density of the phase reference source decreased substantially on the longer VLBA baselines, most probably due to scatter broadening.  When fringe fitting for the phase reference source, no good
solutions were found for the Mauna Kea (MK) and Saint Croix (SC) stations, and all data from these stations had to be discarded. \src was marginally detected, at  a level of 0.25$\pm$0.08\,mJy\,beam$^{-1}$ at the known source position \citep{MiJ11}.

\subsection{Upper limits to the radio quiescent flux}
\label{sec:radiores}
Assuming a radio spectrum of the form $S_\nu\propto\nu^\alpha$, our simultaneous detections in the 4.6 and 7.9 GHz-bands show a spectral index $\alpha$ consistent with $0$ at the 1\,$\sigma$ level for the observations of 2010 May 31. On May 19 the spectral index was consistent with both 0 and -1 at the 2\,$\sigma$ level, while on Apr. 15 $\alpha$ was inconsistent with 0 and consistent with -1 at the 3$\sigma$ level. This can be due to the presence of an optically thin ejection event during the source hard state.  \newline
The EVLA light curve in Figure \ref{licus} shows the fading of the radio counterpart to \src at the end of the outburst. After May 31 (MJD 55347) the source is detected only once, corresponding to  the X--ray re-brightening observed by \chan (see Section \ref{sec:xres}). The most stringent upper limit to the quiescent radio flux  of \src was obtained on 2011 Aug. 2, when the 3$\sigma$ upper limit to the flux density at 8.4 GHz was $<$0.048~$\rmn{mJy\,beam^{-1}}$ (Table \ref{tab:evla}).   

\subsection{The X--ray~--~radio correlation}
\label{sec:LXLR}
\begin{figure*}
\includegraphics[width=15cm, angle=0]{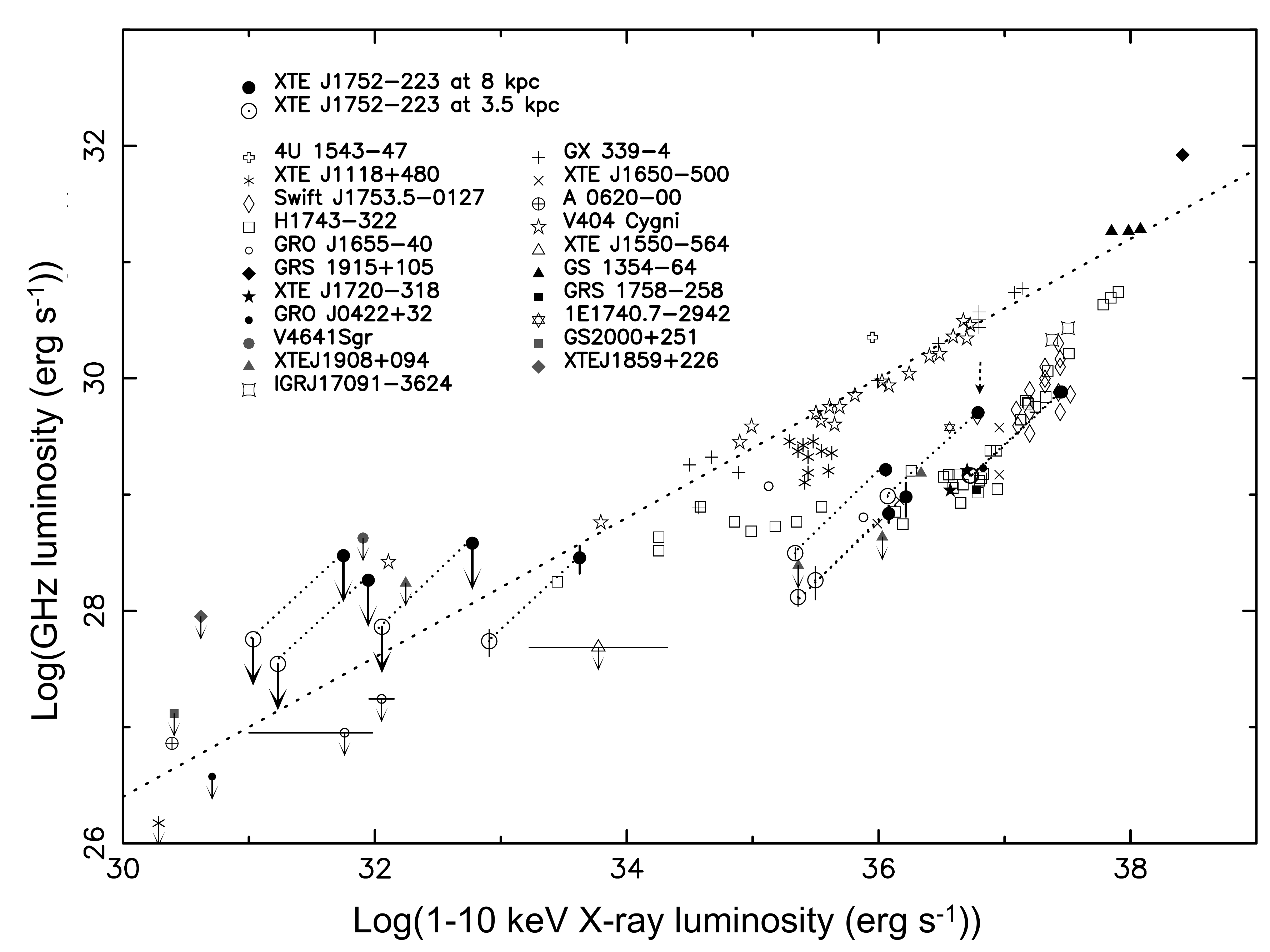} 
\caption{X--ray radio correlation for \src assuming a distance of 8 and 3.5 ${\rmn{kpc}}$ (large black dots and empty circles, connected by dotted lines; see text in Section \ref{sec:LXLR}). The most updated set of sources on the correlation, from \citet{Cal10}, is plotted for comparison. The data for H1743-322 are from a dedicated paper by \citet{Cor11}. Observations of IGR~J17091-3624 in the hard state from \citep{Rod11}, of XTE~J1908$+$094 from \citet{Jon04} and recent radio upper limits from \citet{MiJ11c} are also included, for GRO J0422$+$32, XTE~J1118$+$480, GRO~J1655$-$40, GS2000$+$451, XTE~J1908$+$094, XTE~J1859$+$226 and V4641~Sgr. For clarity we do not show the uncertainty on the source distances, but see \citet{MiJ11c} and \citep{Rod11}. The slope of the standard correlation L$_R\propto$L$_X^{0.6}$ is shown by the dashed line. The data-point indicated by the dashed arrow is from the observation of 2010 Apr. 15, when the radio spectrum was consistent with optically this emission from an ejection event. } 
\label{fig:XR}
\end{figure*}

We performed quasi-simultaneous   ($\lesssim$0.5 days apart)  \chan-EVLA observations on 2010 Jul. 13, 20, 26 and Aug. 2. Moreover, \citet{Rus12} reports \rxte/PCA observations that are less than 0.6 days apart from our EVLA pointings on 2010 Apr. 15, May 19 and May 31 and from our VLBA detection on 2010 Jun. 17. Furthermore, two radio observations of \src during the hard state, performed on 2009 Oct. 31 and Nov. 1 with the Australia Telescope Compact Array (ATCA), are reported by \citet{Bro09}. Quasi-simultaneous \rxte detections \citep{Rus12} provides us with a total of ten X--ray~--Aradio observations (see dashed lines in Figure \ref{licus}) that we can use to investigate the behaviour of \src on the X--ray~--~radio correlation for the hard state of BHTs. As shown by \citet{Jon04}, the non-linearity of the X--ray~--~radio correlation makes the normalization dependent on the distance. In order to compare \src with other sources, we calculated the X--ray (1-10 keV) and GHz radio luminosity assuming a distance of 3.5 kpc as well as a the typical distance of 8 kpc to the Galactic Centre (see discussion on the distance in Section \ref{sec:disc}.). 
When converting the monochromatic radio flux density into a radio luminosity, we multiplied by a frequency of 5 GHz, under the assumption that the spectrum is flat in the GHz range. This assumption leads to an under-estimate of the luminosity for the observation taken on 2010 Apr. 15, when the spectrum was consistent with optically thin synchrotron emission (see Section \ref{sec:radiores} and below). The choice of the 5 GHz frequency has the purpose of comparing with the most updated L$_X$-L$_R$ plot, reported by \citet{Cal10}. Figure \ref{fig:XR} shows the X--ray~--~radio correlation that we obtained for \src together with data from \citet{Cal10} for GX~339$-$4, 4U~1543$-$47, 1E1740.7$-$2942, A~0620$-$00, GS~1354$-$64, XTE~J1118$+$480 and V404~Cygni, all of which follow the `canonical'  X--ray~--~radio correlation, GRS~1915$+$105, which may or may not be an outlier (see \citealt{Cor11}) and the `outliers'  XTE~J1550$-$564, XTE~J1650$-$500, GRO~J1655$-$40, Cygnus~X$-$1, Swift~J1753.5$-$0127, GRO~J0422$+$32, GRS~1758$-$254, XTE~J1720$-$318 and H1743$-$322. For the latter, we have used the hard-state measurements from the work of \citet{Cor11}.  Observations of XTE~J1908$+$094 from \citet{Jon04} and recent radio upper limits from \citet{MiJ11c} for GRO~J0422$+$32, XTE~J1118$+$480, GRO~J1655$-$40, GS~2000$+$451, XTE~J1908$+$094, XTE~J1859$+$226 and V4641~Sgr are also included. 
 \newline
Three of our radio observations did not result in a detection and provide upper limits to the radio flux. The upper limits are consistent with both a standard as well as an under-luminous correlation. On the other hand,  the  six points at  an X--ray luminosity above 10$^{35}$ \ergsec are clearly under-luminous in radio with respect to the standard correlation. The intermediate point at L$_X\sim10^{33}-10^{34}$ \ergsec (depending on the distance) is located much closer to the standard correlation than the higher luminosity ones, resembling the behaviour of H1743$-$322. For a distance of 8 kpc, the  L$_X\sim10^{34}$ \ergsec point falls very well on the standard correlation while the higher luminosity points are located in the region of the (known) `outliers'.  For this distance a transition from the `outliers region' towards the canonical correlation seems to occur around the same luminosity as for  H1743$-$322. Excluding the ATCA detections, for which no uncertainty was reported, a fit to the five remaining detections with a single power-law is very poor ($\chi^2=85$, 3 d.of.) and gives  L$_R\propto$L$_X^{0.51\pm0.04}$. Slightly better fits are obtained with two power-laws, one including the most luminous point and not the faintest, with an index $b=0.87\pm0.08$ ($\chi^2=27$, 2 d.o.f.) and one including the least luminous detection but not the brightest, with $b=0.46\pm0.07$ ($\chi^2=23$, 2 d.o.f.). Still, the fit is poor due to the scatter between the few points. Moreover, the fitted slope of the correlation on the outliers branch relies on the 2010 Apr. 15 observation, for which we are likely under-estimating the radio luminosity and which is not a good representative of the hard state. The fact that the spectral index $\alpha$ was negative indicates contamination from an optically thin ejection event. Despite this, the observations reported by \citet{Bro09} confirm that \src lies on the outliers branch at high luminosity. 



\section{Discussion}
\label{sec:disc}

We have observed the BHT \src towards the end of its 2009-2010 X--ray outburst, with the purpose of exploring the quiescent properties of the source and, in particular, the low-luminosity end of the X--ray~--~radio correlation. To this end we performed four simultaneous EVLA and \chan observations, plus three EVLA observations and one VLBA observation that are simultaneous to RXTE pointings. After a short re-brightening, \src reached a minimum X--ray flux level of (1.6$\pm$0.4)$\times$10$^{-14}$ \ergsec\,cm$^{-2}$ (0.5-10 $\rmn{keV}$) on 2010 Jul. 26. At this same flux level the source was found again one week later, in our last \chan pointing.  
The EVLA observations towards quiescence provide one detection corresponding to the X--ray re-brightening, and allow us to set an upper limit to the quiescent 8.5 GHz radio flux of $<$0.048~$\rmn{mJy\,beam^{-1}}$.\newline
We also observed the source with the Magellan/IMACS instrument in the optical i$^\prime$-band, almost one year after the end of the X--ray outburst. We could not detect the source down to a limiting magnitude of 24.4. 

\subsection{Distance}
With the available data it is not possible to put solid constraints on the source distance. However, some indication of a reasonable distance range can be obtained by combining known empirical relations. 
\citet{Mac03} found that the transition from the hard to soft state at the end of a BHT outburst occurs at a bolometric luminosity that is around the $\sim$2.2\% of the Eddington luminosity L$_{Edd}$ (with an uncertainty of 40\%). 
Assuming a typical BH mass of 10\,\msun we can compute $L_{Edd}$ and, following \citet{Mac03}, the luminosity of \src at the time of the transition from the soft to the hard state. Comparing this luminosity with the RXTE/PCA flux at the time of the transition we can obtain an estimate of the source distance. \citet{Rus12} and \citet{Shap10} show that the RXTE count rate from the source was $~100$ counts per second in the first hard state observation after the soft-to-hard state transition\footnote{As the transition requires several days, during which the source flux varies by a factor of $\sim$2, it is not straightforward to establish what is the luminosity at the transition to be used in the method of \citet{Mac03}. This uncertainty on the time of the transition is included in the 40\% uncertainty on the ratio with the Eddington luminosity indicated by the author. We have tested that, whatever observation we consider during the state transition of \src, we indeed measure consistent distances within 1$\sigma$.}. The corresponding bolometric X--ray flux (for the bolometric correction we follow \citealt{Mac03}) is $\sim2.7\times10^{-9}$ \ergsec\,cm$^{-2}$, from which we obtain a distance of $\sim9.1\pm4.5\,\rmn{kpc}$. This is consistent at the 2$\sigma$ level with the distance of $3.5\pm0.4\,\rmn{kpc}$ claimed by \citet{Shap10} on the basis of the source spectral and timing properties in X--rays. The above methods are both rather uncertain. Their agreement between 3.5 and 9$\,\rmn{kpc}$ suggests that \src is in the Galactic bulge or closer to us, but the boundaries of a likely distance range are hard to define. For this reason, we choose two nominal values:  the 3.5 $\,\rmn{kpc}$ from  \citet{Shap10} (which is probably a conservative lower limit, as will be shown below) and  the distance of 8$\,\rmn{kpc}$ typically assumed for a source in the Galactic bulge. Further in the discussion we will show that a consistent scenario for the multi-wavelength properties of \src emerges within this distance interval, although, given the uncertainties on the relations we used, the extremes of this range are only indicative\footnote{Note that the N$_H$ towards \src (0.5$\times10^{22}\,\rmn{cm}^{-1}$) is close to the Galactic one in the direction of the source(0.45$\times10^{22}\,\rmn{cm}^{-1}$, \citealt{Dic90}). As the Galactic latitude of \src is b=2.1 degrees, the observed N$_H$ implies a lower limit to the source distance that is consistent with the 3.5$\,\rmn{kpc}$ limit derived from the X--ray spectral and timing source properties (assuming a vertical dimension of the dust Galactic plane of $\sim$0.12$\,\rmn{kpc}$, e.g. \citealt{Gre87}).}. 

\subsection{Companion star}
The upper limit we measured on the quiescent  i$^\prime$-band magnitude makes a giant companion star in \src very unlikely, as the source should be located outside of our Galaxy. For $N_H=5\times10^{21}\,{\rmn{cm}^{-2}}$ we obtain an extinction coefficient in the i$^\prime$-band $A_{i\prime}\sim1.54$ (following \citealt{Guv09} and the extinction laws in \citealt{Car89}).  For a G5\,{\sc iii} star the absolute magnitude is $M_{i}\sim-0.27$ (\citealt{Dri2000}, later spectral types are brighter, aggravating the problem). Given our observed magnitude $m_{i}\gtrsim24.4$ and according to the definition of the distance modulus  $5\log d(\rmn{pc})-5 = m_{i}-M_{i}-A_{i}$ we conclude that the distance of a giant would be $d\gtrsim300\,\rmn{kpc}$.  
\newline
An ultra-compact binary scenario, i.e. with a white dwarf donor, is also ruled out by the detection of Hydrogen lines in the optical spectra in outburst \citep{Tor09}. 
Thus, the companion star in \src is most likely a main sequence or sub-giant star. Given our limit on the i$^\prime$-band magnitude, the indications we found on the source distance confine the spectral type of the companion star to a type M or later: a distance of  8\,${\rmn {kpc}}$ gives an absolute magnitude for the secondary $M_{i}\gtrsim7.6$ (reddening has been considered as above) which is true for a main sequence star later than M2 \citep{Dri2000}. If $d=$3.5\,${\rmn {kpc}}$, our i$^\prime$-band non-detection implies that the spectral type has to be later than M5. 

\subsection{Orbital period}
Since \src most likely hosts a main sequence secondary star, we expect the source to follow the relation found by \citet{Sha98} between the V-band outburst amplitude $\Delta$V and the orbital period P$_{orb}$ (note the caveats mentioned by \citealt{MiJ11b} and below). A comparison of the limit to the quiescent i$^\prime$-band magnitude ($\gtrsim24.4$ mag) with observations in outburst indicates an outburst amplitude $\Delta$i$>$8 magnitudes. The corresponding $\Delta$V depends on the spectrum of the disc, which dominates the optical emission in outburst, and on the spectrum of disc$+$companion star in quiescence. We can provide an upper limit to the orbital period  by assuming that the disc is dominating both in outburst and in quiescence. 
\citet{VanP95} found that, on average, B-V$\sim$0 for the accretion disc in a low mass X--ray binary. If V-I is also $\sim$0, $\Delta$V$=\Delta$i$>$8 magnitudes and, according to the correlation of \citet{Sha98}, P$_{orb}\lesssim6.8$\,h. 
As pointed out by  \citet{MiJ11b}, the correlation of \citet{Sha98} does not include the effect of the inclination. High inclination systems (such as MAXI~J1659$-$152) are fainter in the optical during outburst because only a small fraction of the disc surface is visible when the disc is seen edge on. As \src probably hosts a BH, the lack of eclipses or dips still allows for quite a high inclination ($\sim$80 degrees, \citealt{Hor85}). For this reason, although the large outburst amplitude points towards a lower inclination, we can not exclude a significant inclination effect on the \citet{Sha98} correlation. If this is the case, the orbital period of \src would be even shorter that what we are calculating.
\newline
Independent support of our estimate of the distance and the orbital period comes from the correlation between P$_{orb}$ and the luminosity at the peak of an outburst $L_{peak}$ found by \citet{Wu10}. The flux at the peak of the outburst of \src was $\sim8\times10^{-9}$ \ergsec\,cm$^{-2}$ in the 1-10 keV range (\rxte/PCA from \citealt{Rus12}), which gives $0.01$\,L$_{Edd}\lesssim$\,L$_{peak}\lesssim0.05$\,L$_{Edd}$ for a distance $3.5\lesssim d\lesssim8\,\rmn{kpc}$ and a 10\,\msun BH. Using the correlation of \citet{Wu10}, this gives ($0.9\pm0.08)<$\,P$_{orb}<(6.5\pm1.8$)\,h, consistent with the upper limit we derived from the optical outburst amplitude. For $d\sim6\,\rmn{kpc}$, P$_{orb}\sim2.4$\,h, comparable to the shortest orbital period known for a BHT (MAXI~J1659$-$152, \citealt{Kuu11}, \citealt{Ken11}). For 3.5\,$\rmn{kpc}$, the period is much shorter than the shortest known for a BHT, trespassing into the ultra-compact systems regime. The presence of hydrogen in outburst, however, rules out a hydrogen deficient donor. The distance to the source is, therefore, probably larger than 3.5\,$\rmn{kpc}$. The luminosity at the peak of the outburst would also be low at 3.5\,$\rmn{kpc}$ with respect to typical BHTs. Nonetheless, given the uncertainties on the methods we used, we can not rule out such a low distance.
\newline
In summary, the spectral and timing properties of \src and the X--ray luminosity at the transition from soft to hard state indicate a distance of  roughly 3.5$\lesssim$d$\lesssim$8\,${\rmn {kpc}}$, which translates into consistent estimates of the orbital period from two independent methods: the optical outburst amplitude and the X--ray outburst peak luminosity. The methods also suggest that  3.5\,$\rmn{kpc}$ is a conservative lower limit and the source is likely to be further than that. 


\subsection{The X--ray~--~radio correlation}
Using the indications we found on the source distance, we have calculated the X--ray and radio luminosity of \src in order to  compare the X--ray~--~radio correlation for this source with other BHTs. Figure \ref{fig:XR} shows the correlation for  a distance of 3.5 and  8\,${\rmn {kpc}}$. The behaviour of \src resembles that of H1743$-$322, where  a transition  from the region of the `outliers' to the standard correlation occurs when $L_X$ decreases. For d=8\,${\rmn {kpc}}$, \src seems to experience the transition close to the same luminosity where it occurs for H1743$-$322, while for d=3.5\,${\rmn {kpc}}$ it transits at a lower luminosity. The transition was covered with many observations in the case of H1743$-$322 \citep{Cor11} and interpreted as a switch from a radiatively efficient accretion mechanism (on the `outliers' branch) to a radiatively inefficient one (on the `standard' correlation).  It is possible that many BHTs are located on one branch or the other for the full range of luminosity covered within one outburst, or, in other words, that the hard state can be associated with a different accretion flow for different sources. On the other hand, the similarity between \src and H1743$-$322 suggests that the `switching' behaviour of the latter may be shared by other outliers. It is worth noting that, based on our data, it is also possible that \src does not `return' to the L$_R\propto$L$_X^{0.6}$ correlation but crosses it at low luminosities. 
Either way, \src is the first BHT found to show evidence of a transition similar to that of H1743$-$322. More data will be needed in order to confirm this result and to probe the low luminosity end of the X--ray~--~radio correlation better, for \src and for other BHTs. 
 \newline
 
 \subsection{X--ray detection in quiescence?}
The X--ray flux \src on 2010 Jul. 26 and Aug. 2, when we observed it with \chan for the last time, was  (1.4$\pm$0.2)$\times$10$^{-14}$ \ergsec\,cm$^{-2}$ (0.5-10 $\rmn{keV}$, corresponding to a luminosity of  $L_X\sim8\times10^{31}$ \ergsec for a distance of $8\,{\rmn {kpc}}$). 
The fact that we observed consistent flux levels in our last two observations suggests that the source had reached its quiescent level. Nonetheless, it is also possible that we found the source on a temporary plateaux and that it further faded after our observations.  The optical counterpart to \src was far from reaching its quiescent level at the time of our last \chan pointing:  the source flux dropped by more than 3-3.5 magnitudes in the i$^\prime$-band after that moment, until it disappeared below 24.4 mag on 2011 May 3 (see Figure \ref{licus}).  In the case that we measured the quiescent X--ray flux level, this indicates that the decay of the outburst phase towards quiescence had an `inside-out' development, starting with the fading of the X--ray source only later followed by the optical.  Moreover, assuming we measured the quiescent X--ray flux, the quiescent luminosity for our estimate range of distances implies an orbital period longward of 10 hours, if \src follows the trend between orbital period and quiescent X--ray luminosity reported by \citet{Garc01} and \citet{Gal08}. Instead, if the true quiescent luminosity is lower than measured, then the inferred orbital period would be shorter and more compatible with the orbital period inferred above. 

\subsection{X--ray detection of a jet?}
Finally, using our deep \chan observations we discovered unidentified faint sources in the vicinity of \src. In particular, we detected faint X--ray emission at  $\sim$2\farcs9 from the source on 2010 Jul. 26 (source S2 in Figure \ref{chanfinders} and Table \ref{tab:csources}). No clear optical counterpart corresponds to the X--ray position. 
\newline
A possible scenario is that S2 was a transient event aligned by chance with \src. A flare from an unseen background star is unlikely, as the X--ray flux is too high with respect to the optical one. The ratio between X--ray and visual flux for a stellar flare, in fact, is typically Log$(F_X/F_{V})\lesssim-2$. Given the observed X--ray flux from S2, the V-band magnitude of an unseen stellar counterpart should be $\lesssim$18.7, meaning $\lesssim19.2$ in the i$^\prime$-band (V$-$I$=-$0.47 for an O5 star, increasing towards later types, \citealt{Dri2000}). An object with this magnitude would be visible in our deep R-band observation. A chance alignment with an unknown background AGN, binary system or with some peculiar transient event is also rather unlikely due to the rarity of such events, but it can not be ruled out. Nonetheless, the proximity to \src, the position angle (see below) with respect to the BHT core compared to that of the radio jet ejections (\citealt{Yan10}, \citealt{MiJ11b},\citealt{Yan11}), the morphology and the variability of S2 can also be interpreted as X--ray emission coming from a relativistic jet launched by \src. 
\newline Radio observations earlier in the outburst resolved two jet components from \src (\citealt{Yan10}, \citealt{MiJ11}) probably ejected during the outburst at the time of the hard-to-soft transition, in January 2009 \citep{Hom10}. A third one was recently identified by \citet{Yan11}. Besides the radio emission, relativistic jets from BHTs have been found to emit also in the X--rays, at large scales and long after the ejection event (the most extreme case is the BHT XTE J1550-564, \citealt{Corb02}, \citealt{Tom03}, \citealt{Kaa03}). In the BHT H1743-322 \citep{Corb05} \chan observations revealed X--ray emission associated with ejecta previously detected in the radio, at an angular separation of a few arcseconds from the core of the source. Both the radio and X--ray radiation are thought to be synchrotron emission from particles accelerated by shocks within the jet. \newline
The position for S2 from our \chan observation 11055 indicates a position angle of $-52\pm7$ degrees, which is consistent at the 1$\sigma$ level with the position angle of the radio jets,  $50\pm0.6$ degrees (\citealt{MiJ11},  \citealt{Yan11}).
Although not significantly detected by {\sc wavdetect}, fainter emission along the jet direction is visible in observation 11056 too (Figure \ref{chanfinders}). The upper limits to the X--ray flux from S2, obtained from the non-detections in the observation 11056 and in the ones previous to 11055 (see Section \ref{sec:xres}) are high enough that S2 may have had a constant flux level (or possibly re-brightened) between  2010 Jul. 20 (observation 12310) and 2010 Jul. 26 (observation 11055), while it has faded by at least a factor of $\sim$1.8 in the following seven days, until 2010 Aug. 2 (observation 11056).  An X--ray brightening can be caused by a shock in the jet either caused by the collision of consecutive jets traveling at different velocities, by the interaction of the ejecta with the interstellar medium (ISM) or by renewed energization related to the X--ray flare from the core of \src occurred around 2010 Jul. 20 (observation 12310).  \citet{Fen04} proposed a similar scenario for the neutron star X--ray binary Cir~X--1, where  X-ray flares from the source core were causing re-brightening of radio emitting components 2-2\farcs5 downstream in the jets, on a timescale of few days. Although this interpretation has been put into question by recent observations of Cir~X--1\citep{MiJ12}, evidence of a flow of energy through astrophysical jets was found for other objects, such as the NS X--ray binary Sco~X--1 \citep{For01} and several active galactic nuclei (e.g. \citealt{Tin98}). 
For a distance of 8\,$\rmn{kpc}$, and assuming the date of observation 12310 as the starting time (MJD 55397.07034), the velocity of a shock propagating from the core of \src to S2 would be $\gtrsim0.999\,c$ (for details on the calculation see \citealt{Fen04}). Although highly relativistic shocks within the jets were found for other sources (for Sco~X--1, the jets velocity was  0.32-0.57c, but energy appeared to move from the core to the radio lobes at $\gtrsim0.95\,c$, \citealt{For01}) the limit we find for \src is even higher than the extreme case of Cir~X-1, where $\beta\gtrsim0.998\,c$. If \src lies at 3.5\,$\rmn{kpc}$, the velocity would still be $\beta\gtrsim0.994\,c$. Such high values would imply that the jets are very close to the line of sight, with an inclination of less than $\sim$12\degr\,at 3.5\,$\rmn{kpc}$ and less than $\sim$5\degr\,at 8\,$\rmn{kpc}$. As \src did show several X--ray flares during the last part of the outburst, an X--ray and/or radio re-brightening of the source core prior to our \chan observation 12310 could be responsible for the reenergization of S2, leading to smaller velocities for the energy flowing in the jets. A scenario for \src with very low inclination and highly relativistic shocks traveling in the jets  would be consistent with the high proper motion measured from the resolved radio jets ($\sim$58\,$\rmn{masd^{-1}}$, \citealt{Yan11}) and with the fact that no receding jet was detected so far. As pointed out by \citet{Yan11}, \src is a promising Galactic superluminal source candidate.
\newline
Another plausible scenario is that the X--ray emission from S2 is caused by interaction of a previously launched jet with the ISM, or by the collision of two consecutive ejections. Evidence for deceleration of the radio jets launched close to the hard-to-soft transition due to the ISM was already presented by \citet{MiJ11}. The authors found that the motion of the jets was best fit by a combination of a pure ballistic model describing the initial phase after the ejection, followed by a Sedov model further out with respect to the source core.  Extrapolating this model to the time of our detection of S2, the jets should have traveled to a distance of  $\sim$1\arcsec away from the core of \src. This is less than half the separation we observe between \src and S2. If S2 is related to the ejections reported by \citet{Yan10} and \citet{MiJ11}, this result indicates that the jets deceleration did not continue according to the Sedov model due to, e.g., density variations in the ISM. Denser coverage would be needed in order to single out a specific interpretation. At last, we note that the variable, unidentified \chan source S6 also lies on the jet line, albeit on the side opposite S2, at a distance of $\sim$7\farcs4 from the core of  \src. Although it is possible that the X-ray emission we observed from S6 is associated with a receding jet from \src, the source position corresponds to that of a bright star detected in the optical, with a 0.25\% probability of chance coincidence. An association with the optical candidate counterpart is thus likely.

\section{CONCLUSION}

We performed multi-wavelength observations of \src in quiescence and during the last phase of the outburst decay towards quiescence, with the IMACS instrument in the optical i$^\prime$-band, with the \chan satellite in the X--rays, and with the EVLA and VLBA in the radio band. 
We found that the i$^\prime$-band counterpart to the source is fainter than 24.4 magnitudes, while the quiescent radio flux is $<$0.048~$\rmn{mJy\,beam^{-1}}$ at 8.4 GHz. The quiescent X--ray flux as measured from our last \chan observations is (1.4$\pm$0.2)$\times$10$^{-14}$ \ergsec\,cm$^{-2}$ (0.5-10\,$\rmn{keV}$), although we can not rule out a later further dimming of the source. \newline
We presented independent indications that the distance towards \src is likely between $\sim$3.5 and $\sim$8 ${\rmn{kpc}}$, in agreement with previous estimates based on the X--ray spectral and timing properties of the source. We showed that such a distance leads to a coherent picture where   \src has a short orbital period of P$_{orb}\lesssim6.8 {\rmn h}$ and the companion star is later than an M type main sequence star. 
Combining our EVLA pointings  with simultaneous \chan observations and published RXTE data acquired during the outburst, we could investigate the X--ray~--~radio correlation for \src in comparison with other BHTs. We found indications that \src behaves similarly to H1743$-$322, a BHT that is under-luminous in radio with respect to the  `standard' relation L$_R\propto$L$_X^{0.6}$ when above a critical luminosity of $\sim5\times10^{-3}$L$_{Edd}$(M$/$10\msun), but undergoes a transition towards the standard correlation as the luminosity decreases.  This transition was interpreted as a switch from a radiatively efficient accretion mechanism to a radiatively inefficient one. Given that a similar transition occurs in \src, suggests that such changes in the accretion mechanism are not due to some exceptional property of H1743$-$322 but may be shared by other BHTs.
Our deep \chan observations also detected several unidentified X--ray sources in the vicinity of \src, for some of which we found i$^\prime$-band counterparts. One of the X--ray sources is variable and is probably associated with re-energization of jets from \src or with the interaction of the ejecta with the ISM.

\section*{Acknowledgments} \noindent  Mathieu Servillat is acknowledge for providing the CTIO observation of 2010 Aug.~31. EMR acknowledges Sara Motta for the information about the RXTE data. PGJ acknowledges support from a 
VIDI grant from the Netherlands Organisation for Scientific
Research. RW aknowledges support from the European Reasearch council Starting Grants. DS acknowledges a STFC Advanced Fellowship









\bibliographystyle{mn} \bibliography{1752rev.bib}

\end{document}